\documentclass{emulateapj}

\newcommand{\bea}{\begin{eqnarray}}
\newcommand{\eea}{\end{eqnarray}}
\newcommand{\beq}{\begin{equation}}
\newcommand{\eeq}{\end{equation}}

\begin{document}

\def\fun#1#2{\lower3.6pt\vbox{\baselineskip0pt\lineskip.9pt
  \ialign{$\mathsurround=0pt#1\hfil##\hfil$\crcr#2\crcr\sim\crcr}}}
\def\lap{\mathrel{\mathpalette\fun <}}
\def\gap{\mathrel{\mathpalette\fun >}}
\def\kms{{\rm km\ s}^{-1}}
\def\vk{V_{\rm recoil}}

\title{Large Merger Recoils and Spin Flips from Generic Black-Hole Binaries}

\author{Manuela Campanelli\altaffilmark{1},
Carlos Lousto\altaffilmark{2,1},
Yosef Zlochower\altaffilmark{2,1},
David Merritt\altaffilmark{1,3}}
\altaffiltext{1}{Center for Computational Relativity and Gravitation,
School of Mathematical Sciences,
Rochester Institute of Technology, 78 Lomb Memorial Drive, Rochester,
 New York 14623}
\altaffiltext{2}{Center for Gravitational Wave
Astronomy, Department of Physics and Astronomy,
The University of Texas at Brownsville, Brownsville, Texas 78520}
\altaffiltext{3}{Department of Physics,
Rochester Institute of Technology, 85 Lomb Memorial Drive, Rochester,
New York 14623} 

\begin{abstract}

We report the first results from evolutions of generic black-hole
binaries, i.e.\ a binary containing unequal mass black holes with misaligned
spins. Our configuration, which has a mass ratio of $2:1$, consists of an
initially non-spinning hole orbiting a larger, rapidly spinning 
hole (specific spin $a/m = 0.885$), with the spin direction oriented
$-45^\circ$ with respect to the orbital plane. We track the inspiral and merger
for $\sim2$ orbits and find that the remnant receives a substantial kick
of $454\,{\rm km\,s}^{-1}$, more than twice as large as the maximum kick 
from non-spinning binaries. The remnant spin direction is flipped by
$103^\circ$ with respect to the initial spin direction of the larger hole.
We performed a second run with anti-aligned spins, $a/m = \pm0.5$ lying in
the orbital plane that produces a kick of $\sim1830\,{\rm km\,s}^{-1}$
off the
orbital plane. This value scales to nearly $4000\,{\rm km\,s}^{-1}$
for maximally spinning holes.
Such a large recoil velocity opens the possibility that a merged
binary can be ejected even from the nucleus of a massive host galaxy.

\end{abstract}


\maketitle

\section{Introduction}

One of the major goals of numerical relativity has been the accurate
evolution of generic black-hole binaries from inspiral through merger and
ringdown. It is in this non-linear merger regime where most of the
gravitational radiation is emitted, including the
radiation of linear momentum responsible for large merger
recoils that can eject the 
remnant from the host galaxy. 
With the recent breakthroughs in numerical
techniques~\citep{Pretorius:2005gq,Campanelli:2005dd,Baker:2005vv}
this goal is finally being realized.
Within the past 18 months
rapid progress has been achieved in our understanding of
black-hole-binary mergers. Non-spinning equal-mass binaries were
studied in detail, including the last few
orbits~\citep{Campanelli:2006gf,Baker:2006yw}, the effects of
elliptical motion~\citep{Pretorius:2006tp} on the gravitational
radiation, and waveforms generated from binaries with large initial
separations were successfully matched to
post-Newtonian theory with very good agreement~\citep{Buonanno:2006ui,Baker:2006ha,Baker:2006kr}.
Non-spinning unequal mass black holes were studied
in~\cite{Campanelli:2004zw}, \cite{Herrmann:2006ks},
\cite{Baker:2006vn} and \cite{Gonzalez:2006md},
where the recoil velocity of the post-merger remnant was computed.
In particular, the accurate calculations of~\cite{Gonzalez:2006md}
indicate that the maximum recoil velocity of non-spinning
quasi-circular binaries, with mass ratio
$q=m_1/m_2\approx1/3$, is $\sim175\,{\rm km\,s}^{-1}$.
Simulations of highly-spinning black-hole binaries were introduced
in~\cite{Campanelli:2006uy} where it was shown that the direction
of the spin (in that case either aligned or counter-aligned with the
orbital angular momentum) has a strong effect on the  merger time 
and energy-momentum radiated to infinity. In~\cite{Campanelli:2006fg}
it was found that the non-linear
tidal effects were too weak to drive a binary into a corotating state.
Finally in~\cite{Campanelli:2006fy} spin precession and spin-flips were
studied for equal-mass binaries with individual spins not aligned with
the orbital angular momentum (but with individual spins having the
same magnitude and direction).

All the previous simulations contained symmetries which suppressed
some important astrophysical properties (e.g.\ precession, recoil,
spin-orbit coupling) of generic binary mergers, and
in the case of the recoil calculation, spins were neglected entirely.
With the knowledge gained from these simulations we can now design and
evolve a truly prototypical black-hole binary. In the scenario
considered here, a high-mass black hole, with a specific spin of
$a/m=0.885$ (the largest considered thus far), merges with a
smaller hole having negligible spin. The mass ratio of the two
holes is $1.99$, and the initial binary configuration is
such that the spin of the larger hole points $45^\circ$ below
the orbital plane. This configuration will manifest precession of the spin
axis, a significant spin flip of the remnant spin with respect to the
initial individual horizon spin, and a significant recoil kick. The 
simulations that we report in this letter show that the recoil due 
to the spin 
can be more than an order of magnitude larger than the
maximum recoil due to unequal-masses alone.

\section{Techniques}
We use the puncture approach~\citep{Brandt97b} along
with the {\sc TwoPunctures}~\citep{Ansorg:2004ds} thorn to compute
initial data.
  We evolve these black hole binary
data sets using the {\sc LazEv}~\citep{Zlochower:2005bj} implementation
of the moving puncture approach~\citep{Campanelli:2005dd}
(which is based on the BSSN~\citep{Nakamura87,Shibata95, Baumgarte99}
formulation).
We use the Carpet~\citep{Schnetter-etal-03b} mesh refinement driver to provide
a `moving boxes' style mesh refinement. In this approach  refined grids
of fixed size are arranged about the coordinate centers of both holes.
The Carpet code then moves these fine grids about the computational domain
by following the trajectories of the two black holes.
We measure the horizon spin (magnitude and direction) using the techniques
detailed in~\cite{Campanelli:2006fy}. 

\section{Results}

The initial data parameters for our SP6 configuration (i.e.\ generic
binary configuration), which were obtained
using the 3PN equations of motion, are given in
Table~\ref{table:SPNID}. Note that the binary has a small inward radial 
velocity (that we obtain from the post-Newtonian inspiral.) The initial
orbital plane coincides with the $xy$ plane.
\begin{table}
\caption{Initial data parameters for the SP6 (top) and SP2 (bottom)
configurations. $m_p$ is the puncture mass parameter of the two holes.
SP6 has spins $\vec S_1 = (0, S, -S)$ and $\vec S_2 = (0,0,0)$,
momenta $\vec P = \pm (P_r, P_\perp, 0)$, puncture positions $\vec x_1 = (x_+, d, d)$ and
$\vec x_2 = (x_-, d, d)$, and masses $m_1$ and $m_2$.
While SP2 has spins $\vec S_1 = - \vec S_2 = (0,S,0)$, puncture
positions $\vec x_1 = - \vec x_2 = (x,0,0)$, and momenta $\vec P_1 = - \vec P_2 = (0,P,0)$.
}
\begin{tabular}{llllll}
\hline\hline
$m_p/M$ & $0.3185$  & $d/M$   & $0.0012817$ & $m_1/M$  & $0.6680$\\
$x_+/M$ & $2.68773$ & $P_r/M$ & $-0.0013947$ & $m_2/M$  & $0.3355$\\
$x_-/M$ & $-5.20295$ & $P_\perp/M$ &  $0.10695$ & $S/M^2$ & $0.27941$\\
\hline\hline
$m_p/M$ & $0.430213$ & $x/M$ & $3.28413$ & $S/M^2$ & $0.12871$\\
$P/M$ &  $0.13355$ & $m/M$  & $0.5066$ \\
\hline\hline
\end{tabular} \label{table:SPNID} 
\end{table} 

We tested our code with mesh refinement by evolving the SP3 configuration
of~\cite{Campanelli:2006fy}. For this test we evolved SP3 with three
different grid configuration with finest resolutions of 
$M/32$, $M/40$, $M/52$ respectively, and 6 levels of refinement.
We placed the refinement boundaries at the same coordinate distance
from the punctures for each configuration.
We confirmed that the waveforms converge to fourth order and
agree with our unigrid SP3 evolution.

We ran the SP6 run with 7 levels of refinement, with a finest
resolution of $M/43.6$. The outer boundaries were placed at $250M$. We
tracked the individual horizon spins throughout the evolution and
found no significant spin-up of the smaller (initially non-spinning)
hole (the value of $a/m$ at merger was $\sim 10^{-4}$). The larger
black hole, on the other hand, showed a significant $45^\circ$ angle
of spin-precession, with final spin (at merger)
$\vec S_1^{merger}/M^2 = (-0.262, 0.189, - 0.214)$. During the evolution
the binary performed $\sim1.8$
orbits prior to the formation of the common apparent horizon (CAH).
The first common
horizon was detected at $t_{\rm CAH}/M=197.96\pm0.07$.  The common
horizon had mass $M_{\cal H}/M = 0.9781\pm0.0001$ indicating that
$(2.19\pm0.01)\%$
of the mass was converted into gravitational radiation.  The spin of
the remnant horizon was $\vec S_{\rm rem}/M^2 = (-0.0397\pm0.0005,
0.242\pm0.002, 0.4097\pm0.0002)$. The ADM angular momentum for this
system is $\vec J_{\rm ADM}/M^2 = (0, 0.27941, 0.56447)$, thus we predict that
the radiated angular momentum is $\vec J_{rad}/M^2 =
(0.0397\pm0.0005,0.037\pm0.002, 0.1548\pm0.0002)$.
The measured radiated mass and
angular momentum, based on the $\ell=2$ through $\ell=4$ modes of
$\psi_4$ were $E_{\rm rad}/M = 0.0218\pm0.0004$ and
 $\vec J_{rad} = (0.04\pm0.01,
0.04\pm0.01, 0.16\pm0.01)$, which agree well with the remnant horizon parameters.
Note that the agreement in $\vec J_{rad}$ between the horizon spin and
radiation calculation indicates that our method for calculating the
spin direction is reasonably accurate for our choice of coordinates. The final
horizon spin is flipped by $103^\circ$ with respect to the initial
spin of the larger individual horizon and $33^\circ$ with respect to the
initial orbital angular momentum. 

\begin{figure}
\begin{center}
\includegraphics[width=2.2in]{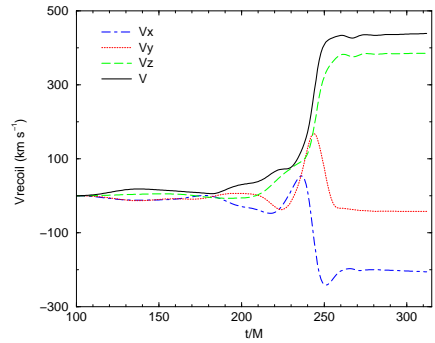}
\caption{The recoil velocities for the SP6 configurations as measured for
an observed at $r=30M$. 
}
\label{fig:recoilvelocities}
\end{center}
\end{figure}

The remnant hole acquires a significant recoil velocity of $\vec
\vk = (-208\pm30, -48\pm7, 424\pm10) {\rm km\, s}^{-1}$
(see Fig.~\ref{fig:recoilvelocities}),
which makes an angle of
$27^\circ$ with respect to the initial orbital angular momentum and
$135^\circ$ with respect to the initial spin.
We measured this kick by calculating the radiated linear momentum
~\citep{Campanelli99} based on the $\ell=2$ through $\ell=4$ modes
of $\psi_4$.
We extracted these modes at $r=25M, 30M, 35M, 40M$ and then
extrapolated the radiated momenta
calculated at these radii to $r=\infty$ using a linear (least-squares)
fit (we excluded the initial data
burst from this momentum calculation). The quoted errors in $\vec V_{\rm
recoil}$ are the differences between the linear extrapolation and 
a quadratic extrapolation. This recoil velocity of
$454\pm 25\,{\rm km\, s}^{-1}$ is more than double the maximum 
recoil velocity found for non-spinning
holes~\citep{Gonzalez:2006md} even including small eccentricity effects
\citep{Sopuerta:2006et}.
Furthermore, the spin-induced recoil in the $xy$ plane might be offset
by the mass-difference-induced recoil, potentially implying that a
rotation of the spin about the $z$-axis may lead to 
a significantly larger in-plane component of the recoil velocity.
Further study will be needed to determine if the
mass-difference-induced recoil is (partially) aligned or counter-aligned
with the spin-induced recoil.

\section{Discussion}

We studied, for the first time using fully non-linear numerical
relativity, a realistic astrophysical configuration of unequal mass,
spinning black holes starting from a slightly elliptical orbit, with
radial inward velocity as predicted by post-Newtonian theory, for
large initial separations. Our main new result is that
the spin component to the recoil velocity may produce the leading contribution.
This is suggested by the fact that the $z$-component of the recoil velocity,
which is not present for non-spinning binaries,
is the dominant component.  This also can be seen from 
the 2nd post-Newtonian expressions for the radiated linear momentum~\citep{Kidder:1995zr}
\begin{eqnarray}\label{PNPdot}
{\dot {\vec P}} &=& - {8 \over 15} {\mu^2 m \over r^5}
\Bigl\{ 4 \dot r ({\vec v \times \vec\Delta})
- 2v^2 ({ \hat n \times \vec\Delta}) \nonumber \\ && \mbox{}
- ({ \hat n \times v}) \left[ 3\dot r ({ \hat n \cdot \vec\Delta})
+ 2 ({\vec v \cdot \vec\Delta}) \right] \Bigr\},
\end{eqnarray}
 where ${\vec x} \equiv {\vec x_1}-{\vec x_2}$, $r \equiv |\vec x|$, ${\vec v}={d{\vec x}/dt}$,
${ \hat n}\equiv{{\vec x}/r}$, $\mu \equiv m_1m_2/m$,
$m=m_1+m_2$,
${\vec \Delta} \equiv m({\vec S_2}/m_2 -{\vec S_1}/m_1)$, and an overdot
denotes $d/dt$.

Based on this expression we can predict that the maximum recoil
velocity is reached for equal mass black holes with opposite
(and maximal) spins lying on the orbital plane since all four terms add
constructively to the radiated momentum.
We performed one additional run, denoted by SP2,
with anti-aligned spins of magnitude
$a/m\pm0.5$ lying initially along the y-axis as reported in 
Table~\ref{table:SPNID}. We obtain a 
$\vec\vk = (0,0,1830\pm 30) {\rm km\, s}^{-1}$. By rescaling this to maximal
spins we obtain essentially double those values raising the maximum recoil 
of spinning holes to almost $4000\,{\rm km\, s}^{-1}$.

Equation~(\ref{PNPdot})
also allows us to propose an empirical formula for the
total recoil velocities
\footnote{After completion of this work we become aware of a new paper \cite{Baker:2007gi} also modeling $v_\perp$}
\begin{eqnarray}\label{empirical}
\vec{V}_{\rm recoil}(q,\vec\alpha)&=&v_m\,\hat{e}_1+
v_\perp(\cos\xi\,\hat{e}_1+\sin\xi\,\hat{e}_2)+v_\|\,\hat{e}_z,\nonumber\\
v_m&=&A\frac{q^2(1-q)}{(1+q)^5}\left(1+B\,\frac{q}{(1+q)^2}\right),\nonumber\\
v_\perp&=&H\frac{q^2}{(1+q)^5}\left(\alpha_2^\|-q\alpha_1^\|\right),\nonumber\\
v_\|&=&K\cos(\Theta-\Theta_0)\frac{q^2}{(1+q)^5}\left(\alpha_2^\perp-q\alpha_1^\perp\right),
\end{eqnarray}
where $\vec{\alpha}_i=\vec{S}_i/m_i^2$, the index $\perp$ and $\|$ refer to
perpendicular and parallel to the orbital angular momentum respectively,
$\hat{e}_1,\hat{e}_2$ are
orthogonal unit vectors in the orbital plane and $\xi$ measures the
angle between the unequal mass and spin contribution to the recoil
velocity in the orbital plane. The constants $H$ and $K$ can be determined
from newly available runs. The angle $\Theta$ is defined as the angle 
between the in-plane component of $\vec \Delta$ and the infall direction
at merger.
 We have confirmed this $\cos\Theta$ dependence
by evolving a set of runs similar to SP2 but with initial spins
rotated by $\delta\Theta = \pi/4$, $\pi/2$, and $\pi$. The resulting
kicks were well modeled by a $\cos(\Theta - \Theta_0)$ dependence,
with $V_z = (1873\pm30) {\rm km\, s^{-1}} \cos[\delta \Theta -
(0.18\pm0.02)]$. Note that we measured a maximum kick of
$(1830\pm30) {\rm km\,s^{-1}}$ for $\delta \Theta=0$ and 
$\delta \Theta = \pi$,
and a minimum kick of $(352\pm10){\rm km\,s^{-1}}$ for $\delta \Theta =-
\pi/2$. This will be the subject of an upcoming paper by the
authors.
The total recoil velocity also acquires a
correction~\citep{Sopuerta:2006et} for small eccentricities, $e$, of
the form $\vec{V}_{e}=\vec{V}_{\rm recoil}\,(1+e)$.

From~\cite{Gonzalez:2006md} $A=1.2\times10^4\,{\rm km\, s}^{-1}$ and $B=-0.93$.
From fits to \cite{Herrmann:2007ac} and \cite{Koppitz:2007ev} we find 
$H=(7.3\pm0.3)\times10^3\,{\rm km\, s}^{-1}$ and from SP2 and
~\cite{Gonzalez:2007hi} (which appeared after this paper had been
submitted)
we find $K\cos(\Theta-\Theta_0)=(6,-5.3)\times10^4\,{\rm km\, s}^{-1}$ respectively. Note
the sign difference showing some of the $\vk$ dependence on $\Theta$.

The in-plane recoil velocity for our SP6 configuration is consistent
with Eq.~(\ref{empirical}) with $\xi\approx88^\circ$,
however our simulation shows strong precession of
the spin near merger, where a large fraction of the recoil velocity is built up,
hence it is difficult to accurately determine the spin parameters $\vec\alpha$
and $\Theta$ to be used in Eq.~(\ref{empirical}).

\section{Astrophysical Implications}

A number of arguments \citep{Shapiro05,Gammie04} suggest that spins
of supermassive black holes (SMBHs) are close to maximal, $a/m\gap 0.8$,
and perhaps as great as $0.99$ \citep{Reynolds05}.
Mass ratios of binary SMBHs are poorly constrained observationally,
but the luminous galaxies known to harbor SMBHs are believed to have
experienced a few to several major mergers (mass ratios
$0.3\lap q \lap 1$) over their lifetimes \citep{Haehnelt02,Merritt06};
mergers with $q\approx 1$ were more common in the past.
Together with the results discussed above, these arguments suggest that
recoil velocities for binary SMBHs in galactic nuclei are often
of order  $\sim 10^3\ \kms$.
Here, we consider some consequences of such large recoil velocities.

{\it Ejection:} Central escape velocities from giant elliptical galaxies
and spiral galaxy bulges
are $450\ {\rm km\ s}^{-1} \lap v_e\lap 2000\ {\rm km\ s}^{-1}$,
dropping to $\lap 300\ \kms$ in dwarf elliptical (dE) and dwarf 
spheroidal (dSph) galaxies \citep{Merritt:2004xa}.
Recoil velocities as large as $10^3\ {\rm km\ s}^{-1}$
would easily eject SMBHs from dE and dSph galaxies, 
and in fact there is little evidence
for SMBHs in these galaxies.
However, we note that the mass dependence of spin-dominated kicks,
$V_{\rm recoil}\sim q^2$, implies that recoil velocities might
only infrequently be as large as in the equal-mass case.
If the tight empirical relations between SMBH mass and luminous
galaxy properties \citep{FM00,Marconi03}
are to be maintained,
peak recoil velocities are constrained to be $\lap 500\ \kms$
\citep{Libeskind06};
the upper limit on $\vk$ is relaxed if most of the
merger-induced growth of SMBHs took place at low redshifts ($z\lap 2$)
when potential wells were deeper \citep{Libeskind06}.
Ejection of SMBHs from shallow potential wells
at high redshift implies a maximum $z$ at which the progenitors of
present-day SMBHs could have started merging \citep{Merritt:2004xa},
and the existence of bright quasars at $z\approx 6$ is difficult
to reconcile with recoil velocities $\gap 10^2\ \kms$ unless
their SMBHs grew very quickly via accretion \citep{Haiman04}.
However we note that these and similar conclusions are based on an assumed 
mass ratio dependence for the kicks that is invalid if 
recoil velocities are dominated by spin effects.

{\it Displacement:} The long return time for a SMBH ejected near
escape velocity implies a substantial probability of finding
a displaced SMBH in a luminous E galaxy, especially if the latter
was the site of a recent  merger \citep{Merritt:2004xa,Madau04,Vicari06}.
A recoiling SMBH carries with it material that was orbiting
with velocity $v\gap \vk$ before the kick;
the size of the region containing this mass is
\begin{equation}
{GM_\bullet\over \vk^2}
\approx 1\ {\rm pc}\ M_8 \sigma^{-2}_{200}
\left({v_e\over \vk}\right)^{2}
\label{eq:reff}
\end{equation}
where $M_8$ is the SMBH mass in units of $10^8 M_\odot$ and
$\sigma_{200}$ is the nuclear velocity dispersion in units
of $200\ \kms$.
This radius is sufficient to include the
inner accretion disk and the
broad-line region gas, implying that a kicked BH
can continue shining for some time as a quasar.
Plausibility of models that explain the ``naked'' quasar
HE 0450-2958 as an ejected SMBH \citep{Haehnelt06} are
enhanced by the larger kick velocities found here,
since $v_e$ from the nearby galaxy is $\gap 500\ \kms$;
however the presence of spectral features associated
with narrow emission line region gas in the quasar
is still difficult to reconcile with the recoil hypothesis
\citep{Naked06}.

{\it Galaxy cores:} The kinetic energy of a displaced SMBH
is transferred to the stars in a galactic nucleus via
dynamical friction, lowering the stellar density and
enlarging the core before the hole returns to its central
location \citep{Merritt:2004xa,Boylan04}.
``Damage'' to the core is maximized for $\vk/v_e\approx 0.7$
\citep{Merritt:2004xa}, hence the effect could be large even
in the brightest E galaxies.
Observed core masses are mostly in the range $0.5-1.5 M_\bullet$,
consistent with the cores having been generated by binary SMBHs
without  the help of kicks \citep{Merritt06}; however a significant
fraction have core masses exceeding $2M_\bullet$ suggesting
an additional contribution from recoils.
Anomalously  large cores in BCG's (``brightest cluster galaxies'')
might be explained in this way, since these galaxies have
experienced the largest number of mergers; this explanation
would lessen the necessity for alternatives that
require BCGs to contain hypermassive BHs \citep{Lauer06}.

{\it Jet Directions:} We measure both a significant pre-merger spin precession of $\sim 45^\circ$ and a
change in excess of $90^\circ$ between the initial and final spin vectors, verifying
the spin-flip phenomenon first discussed by \cite{Merritt:2002hc}.  Thus, our
simulation represents a possible model for the 
merger process responsible for generating radio sources with
changing jet directions.
In particular, the highly-spinning large
mass black hole merging with the smaller mass non-spinning hole is a possible
model for both the gradual semi-periodic deviations of
the jet directions from a straight line~\citep{2006MmSAI..77..733K}
due to precession of the spin (and hence jet) direction,
and the abrupt change in jet direction forming X-shaped
patterns~\citep{1985A&AS...59..511P,1992ersf.meet..307L}.

\acknowledgments 
We thank Erik Schnetter for valuable discussions and providing {\sc Carpet}.
We thank Marcus Ansorg for providing the {\sc TwoPunctures} initial data thorn
and Johnathan Thornburg for providing {\sc AHFinderDirect}. 
We gratefully acknowledge 
NSF for financial support from grant PHY-0722315.
D. M. was supported by grants AST-0420920 and AST-0437519 from the NSF and
grant NNG04GJ48G from NASA.
Computational resources were provided by
Lonestar cluster at TACC.
M.C. thanks the Center for Gravitational Wave Astronomy, University
of Texas, Brownsville for travel
and computing support.

\end{document}